\documentclass[12pt]{iopart}
\usepackage{times}
\usepackage{graphicx}

\begin{document}
\title{Normal Order: Combinatorial Graphs}
\author{
Allan I. Solomon$^{\dag\S}$,
Gerard Duchamp$^{\S\S}$,
Pawel Blasiak$^{\dag\ddag}$,
Andrzej Horzela$^{\ddag}$,
Karol A. Penson$^{\dag}$ }
\address{
\vspace{10pt}$^{\dag}$Universit\'{e} Pierre et Marie Curie\\
Laboratoire   de  Physique   Th\'{e}orique  des  Liquides, CNRS UMR 7600\\
Tour 16, $5^{i\grave{e}me}$ \'{e}tage, 4, place Jussieu, F 75252
Paris, Cedex 05, France\\ \vspace{5pt}
$^{\S\S}$LIFAR, Universit\'{e} de Rouen\\
76821 Mont-Saint Aignan Cedex, France\\ \vspace{5pt}
$^{\ddag}$H. Niewodnicza{\'n}ski Institute of Nuclear
Physics, Polish
Academy of Sciences\\
Department of Theoretical Physics\\
ul. Radzikowskiego 152, PL 31-342 Krak{\'o}w, Poland
\\\vspace{5pt}
$^\S$The Open University\\
Physics and Astronomy Department\\
Milton Keynes MK7 6AA }
\begin{abstract}
A conventional context for supersymmetric problems arises when we consider systems containing both boson and
fermion operators. In this note we consider the normal ordering problem for a string of such operators. In the
general case, upon which we touch briefly, this problem leads to combinatorial numbers, the so-called {\em Rook
numbers}. Since we assume that the two species, bosons and fermions, commute, we subsequently  restrict ourselves
to consideration of a single species, single-mode boson monomials. This problem leads to elegant generalisations
of well-known combinatorial numbers, specifically  Bell and Stirling numbers. We explicitly give the generating
functions for some classes of these numbers. In this note we concentrate on the  combinatorial  graph approach,
showing how some important classical results of graph theory lead to transparent representations of the combinatorial
numbers associated with the boson normal ordering problem.
\end{abstract}


\section{Normal Ordering}
In this note we give a brief review of some combinatorial graphs associated with the normal ordering of creation
and annihilation operators.

The process of normally ordering a string of creation and annihilation operators simply means reordering the
elements of the string so that all the annihilation operators appear on the right, taking into account the
commutation (or anti-commutation) relations. The value of such a procedure is that, for example, the expectation
value in a coherent state of such a normally ordered string may be immediately seen. Performing this operation on
strings of bosons or fermions leads to classical combinatorial numbers and, in the analogous {\em quon} case, to
$q$-variants of these numbers. In a ``supersymmetric'' context one may well have to consider a general string of
bosons and fermions; but as the bosons and fermions are assumed to commute, the ordering reduces trivially to
ordering the two species separately. We thus assume the basic commutation relations
\begin{itemize}
\item Bosons $[a_i,{a_j}^\dag]={\delta}_{ij}$
\item Fermions $\{f_i,{f_j}^\dag\}={\delta}_{ij}$
\item Mixed $[a_i,{f_j}^\dag]=0$
\end{itemize}
We note that the boson ordering problem is equivalent to ordering
a string of operators $a\equiv d/dx$ and $a^{\dagger}\equiv x$.

For both bosons and fermions, the ordering process of a general string results in combinatorial numbers called
{\em Rook numbers}. We shall merely give a brief account of this approach here; further details, especially with
respect to the fermion case, may be found in Ref. \cite{nav}. In this note we {\em illustrate} rather than
{\em prove} the methods described.
\section{Rook numbers and generalized Stirling numbers}
The normal form of a bosonic string or ``word'', $w=w(a,a^\dag)$, satisfies
$\mathcal{N}(w)=w$. Normal forms of boson strings are connected to the so-called {\em
Rook numbers} \cite{B}, \cite{St1} in the following way \cite{Leeuwen}. Draw a
North-East ($\nearrow$) line over each creation operator $a^{\dagger}$ and a
South-East ($\searrow$) line over each annihilation operator $a$ so as to produce a
continuous line. This gives the diagram  below.

\begin{center}
\setlength{\unitlength}{1mm}
\begin{picture}(80,60)
\thinlines \linethickness{0.075mm} \multiput(0,0)(8,0){11}{\line(0,1){56}} \multiput(0,0)(0,8){8}{\line(1,0){80}}
\linethickness{0.4mm} \put(0,0){\line(1,1){8}} \put(8,8){\line(1,1){8}} \put(16,16){\line(1,-1){8}}
\put(24,8){\line(1,1){8}} \put(32,16){\line(1,-1){8}} \put(40,8){\line(1,1){8}} \put(48,16){\line(1,1){8}}
\put(56,24){\line(1,1){8}} \put(64,32){\line(1,1){8}} \put(72,40){\line(1,-1){8}} \put(0,0){\line(1,1){8}}
\put(2,-5){$a^{\dagger}$} \put(10,-5){$a^{\dagger}$} \put(19,-5){$a$} \put(26,-5){$a^{\dagger}$} \put(35,-5){$a$}
\put(42,-5){$a^{\dagger}$} \put(50,-5){$a^{\dagger}$} \put(58,-5){$a^{\dagger}$} \put(66,-5){$a^{\dagger}$}
\put(75,-5){$a$} \multiput(16,16)(1,1){40}{\circle*{0.1}} \multiput(56,56)(1,-1){16}{\circle*{0.1}}
\multiput(32,16)(1,1){32}{\circle*{0.1}} \multiput(24,24)(1,-1){8}{\circle*{0.1}}
\multiput(32,32)(1,-1){16}{\circle*{0.1}} \multiput(40,40)(1,-1){16}{\circle*{0.1}}
\multiput(48,48)(1,-1){16}{\circle*{0.1}}
\end{picture}
\label{figr1}
\end{center}
\begin{center}
\bigskip
\end{center}
 \bigskip \bigskip
The picture is completed by the dotted lines as shown. The result of this process is a
{\em Ferrers diagram} or {\em board} \cite{St1} representing, in this case, the
partition $\{5,4\}$ of $9$ represented by the following  diagram.

\begin{center}
\setlength{\unitlength}{1mm}
\begin{picture}(70,30)
\multiput(0,0)(0,10){2}{\line(1,0){50}}
\multiput(0,0)(10,0){6}{\line(0,1){10}}
\put(0,20){\line(1,0){40}}
\multiput(0,10)(10,0){5}{\line(0,1){10}}
\end{picture}
\end{center}
\begin{center}
 \end{center}

The $k$-th {\it rook number} $r_k(B)$ of a Ferrers board $B$  is the number of ways of
placing $k$ non-capturing rooks on the board. For our example of the partition
$\{5,4\}$ one has

\begin{center}
\begin{tabular}{c|c|c|c|c}
$k$     & 0     & 1 & 2     & $k>2$\\
\hline
$r_k(B)$    & 1 & 9 & 16    & 0
\end{tabular}
\end{center}
and thus the normal form of $w=a^{\dagger}a^{\dagger}aa^{\dagger}aa^{\dagger}a^{\dagger}a^{\dagger}a^{\dagger}a$
is given by
\begin{equation}\label{1}
\mathcal{N}(w)=a^{\dagger}a^{\dagger}a^{\dagger}a^{\dagger}a^{\dagger}a^{\dagger}a^{\dagger}aaa+9\
a^{\dagger}a^{\dagger}a^{\dagger}a^{\dagger}a^{\dagger}a^{\dagger}aa+16\
a^{\dagger}a^{\dagger}a^{\dagger}a^{\dagger}a^{\dagger}a .
\end{equation}
For $w=aa^{\dagger}a^2(a^{\dagger})^2$, the board represents the partition $[3,2,2]$ and
\begin{eqnarray}\nonumber
\mathcal{N}(aa^{\dagger}a^2(a^{\dagger})^2)
&=&a^{\dagger}a^{\dagger}a^{\dagger}aaa+7\ a^{\dagger}a^{\dagger}aa+10\ a^{\dagger}a+2\\\label{2}
&=&r_0(B)(a^{\dagger})^3a^3+r_1(B)(a^{\dagger})^2a^2+r_2(B)(a^{\dagger})a+r_3(B).
\end{eqnarray}
In fact Eqs.(\ref{1}) and (\ref{2}) are illustrations of the
general formula expressing the normal form $\mathcal{N}(w)$ with the
help of the rook numbers as
\begin{equation}\label{3}
\mathcal{N}(w)=\sum_{k=0}^\infty r_k(B):w^{(k)}:
\end{equation}
In Eq.(\ref{3}) $:w^{(k)}:$ means that in the word $w=w(a,a^\dag)$ we cross out
$k$ $a$'s and $k$ $a^\dag$'s and then order normally the result {\em without}
taking into account the commutation relations. In fact for finite word
$w$ the sum in Eq.(\ref{3}) has a finite number of non-vanishing terms.
Note that coefficients of the normal monomials are the rook numbers of the board; it is possible to give a simple
algorithm which computes these numbers.

 Similar observations apply to the normal ordering of {\em quons} $a_q$
(q-bosons) satisfying $$[a_q,a_{q}^\dag]_q\equiv a_qa_{q}^\dag-qa_{q}^\dag a_q=1,$$ the $q$-Weyl
algebra\footnote[1]{An interesting, if as yet experimental, observation is that the resulting polynomials in $q$
obtained by setting $a_q=1=a_{q}^{\dagger}$ are {\em unimodal}.}.

If we restrict ourselves to simple recurring strings in one boson mode, which we shall henceforth do, we see that
classical combinatorial numbers  appear naturally \cite{BPS1}, \cite{AoC}.

The normal ordering problem for canonical bosons  $[a,a^\dag]=1$ is
related to certain combinatorial numbers $S(n,k)$ called {\em Stirling
numbers of the second kind} through \cite{Katriel1}
\begin{eqnarray}
(a^\dag a)^n=\sum_{k=1}^nS(n,k) (a^\dag)^k a^k,
\end{eqnarray}
with corresponding numbers $B(n)=\sum_{k=1}^n S(n,k)$ called {\em Bell numbers}. In fact, for physicists, these
equations may be taken as the {\em definitions} of the Stirling and Bell numbers.  For the quons $a_q$ (q-bosons)
mentioned above, a natural $q$-generalisation \cite{Katriel2}, \cite{Schork} of these numbers is
\begin{eqnarray}
(a_{q}^\dag a_q)^n=\sum_{k=1}^nS_{q}(n,k) (a_{q}^\dag)^k a_{q}^k.
\end{eqnarray}
In the {\em canonical} boson case, for integers
$n,r,s>0$ we define generalized Stirling numbers of the second
kind $S_{r,s}(n,k)$ through ($r\geq s$):
\begin{eqnarray}\label{C}
[(a^\dag)^ra^s]^n=(a^\dag)^{n(r-s)}\sum_{k=s}^{ns}S_{r,s}(n,k)(a^\dag)^ka^k,
\end{eqnarray}
as well as generalized Bell numbers $B_{r,s}(n)$
\begin{eqnarray}\label{D}
B_{r,s}(n)=\sum_{k=s}^{ns}S_{r,s}(n,k).
\end{eqnarray}
For both $S_{r,s}(n,k)$ and $B_{r,s}(n)$ exact and explicit formulas
have been found \cite{BPS1}, \cite{AoC}.  We refer the interested reader to these
sources for further information on those extensions.
However, in this note we shall mainly deal with the classical Bell
and Stirling numbers, corresponding to $B_{1,1}(n)$ and $S_{1,1}(n)$ in our notation, and the extension to $B_{2,1}(n)$
and $S_{2,1}(n)$ .

The conventional and picturesque description of the classical Bell and Stirling Numbers of the second kind is in
terms of the distribution of differently coloured balls among identical containers or, equivalently, the number of
partitions of an $n$-element set \cite{slo}, \cite{Comtet}. The relation of this classical definition to the normal order
expansion of $(a^{\dagger}a)^n$ is via the contractions induced by the application of Wick's Theorem.  In general,
$S(n,m)$ gives the coefficient of the term $:(a^{\dagger}a)^m:\equiv (a^\dagger)^m a^m$ in the normal ordering
expansion, directly from the  (physicist's) definition of $S(n,m)$.

Taking the concrete example $(a^{\dagger}a)^3$,
\begin{itemize}
\item $S(3,1)=1$ is the coefficient of the term $:a^\dagger a:\equiv a^\dagger a$
\item $S(3,2)=3$ is the coefficient of the term $:(a^\dagger a)^2:\equiv (a^\dagger)^2 a^2$
\item $S(3,3)=1$ is the coefficient of the term $:(a^\dagger a)^3:\equiv (a^\dagger)^3 a^3$
\end{itemize}
In the next section we relate these combinatorial numbers to certain graphs. We shall elaborate on this  graphical
approach elsewhere \cite{tobepub}.
\section{Generating Functions and Graphs}
In general, for combinatorial numbers $g(n)$ we may define an {\em
exponential generating function} $G(x)$ through \cite{Wilf}
\begin{equation}
G(x) = \sum_{n=0}^{\infty} g(n)\frac{x^n}{n!}.
\end{equation}
For the Bell numbers, this generating function takes the particularly
nice form \cite{Comtet}
\begin{equation}
G(x) = \sum_{n=0}^{\infty} B(n)\frac{x^n}{n!}=\exp(\exp(x)-1).
\end{equation}
We shall derive this exponential generating function by the use of a simple graph  theory technique.
Some initial terms of the sequence $\{B(n)\}$ are $\{1,2,5,15,52,203,877,\ldots \}$.

Another convenient way of representing combinatorial numbers is by means of {\em graphs}. To illustrate this, we
now consider a graphical method for describing the combinatorial numbers $B(n)$ associated with the normal order
expansion of $(a^{\dagger}a)^n$. For concreteness, take the cases $n=1,2,3$ (Figure 1). In this diagram the filled
dots on the left represent the differently coloured balls, while the empty dots represent the identical
containers.  This pictorial approach  gives the values for
$S(n,m)$. By convention, $S(n,0)=\delta_{n,0}$.
\begin{figure}
\vspace{1cm}
\begin{center}\resizebox{10cm}{!}{\includegraphics{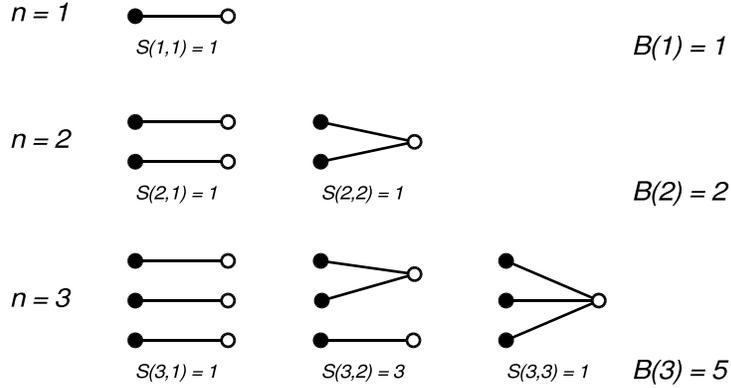}}
\caption{Graphs for $B(n),\; \; \; n=1,2,3.$}
\end{center}
\end{figure}

One reason the graphical representation is useful is that there exist some rather powerful results which apply to
graphs and their associated generating functions. We list and apply two such results, labelled A and B, in what
follows.\\ \\
{\bf A. Connected graph theorem}\\ \\
This  states that if $C(x)=\sum_{n=1}^{\infty} c(n)x^n/n!$ is the exponential generating function of {\em labelled
connected} graphs, {\em viz.} $c(n)$ counts the number of connected graphs of order $n$, then
\begin{equation}
A(x) = \exp(C(x))
\end{equation}
is the exponential generating function for {\em all} graphs.

We may apply this very simply to the case of the $B(n)$  graphs in  Figure 1.
For each order $n$, the {\em connected} graphs clearly consist of a single graph.
Therefore for each $n$ we have $c(n)=1$; whence, $C(x)=exp(x)-1.$
It follows that the generating function for {\em all} the  graphs $A(x)$ is given by
\begin{equation}
A(x)=\exp(\exp(x)-1)
\label{bell}
\end{equation}
which is the generating function for the Bell numbers.

Such graphs  may be generalised to give  graphical representations for the extensions $B_{r,s}(n)$ \cite{tobepub}.
We illustrate this by using the following powerful result on certain classes of graphs:\\   \\
{\bf B. Generating function for a class of graphs} \\ \\
We generalize the graphical representation for the Bell numbers given in Figure 1. As
before, we shall be counting labelled  lines. A line starts from a black dot, the {\em
origin}, and ends at a white dot, the {\em vertex}. What we refer to as {\em origin}
and {\em vertex} is, of course, arbitrary. At this point there are no other rules,
although we are at liberty to impose further restrictions; a black dot may be the
origin of 1,2,3,... lines, and a white dot the vertex for 1,2,3,... lines. We may
further associate {\em strengths} $V_s$ with each vertex receiving $s$ lines, and
multipliers $L_m$ with a black dot which is the origin of $m$ lines. Again  $\{V_s\}$
and $\{L_m\}$ play symmetric roles; in this note we shall only consider cases where
the $L_m$ are either $0$ or $1$.

\begin{figure}
\vspace{1cm}
\begin{center}\resizebox{10cm}{!}{\includegraphics{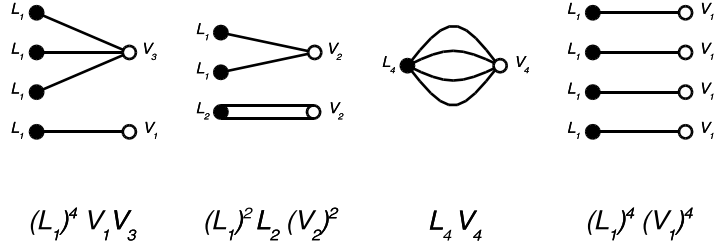}}
\caption{Some examples of 4-line graphs.}\label{FigC}
\end{center}
\end{figure}

In Figure \ref{FigC} we illustrate these rules for four different graphs corresponding to $n=4$.

There is an exponential generating function $G(x,V,L)$ which counts the number $g(n)$ of graphs with $n$ lines
arising from the above rules \cite{Vasiliev}, \cite{ben1}:
\begin{eqnarray}
G(x,V,L)&=&\left.\exp\left(\sum_{m=1}^{\infty} L_m \frac{x^m}{m!}\frac{d^m}{dy^m}\right)
 \exp\left(\sum_{s=1}^{\infty} V_s \frac{y^s}{s!}\right)\right|_{y=0} \nonumber\\
&\equiv&\sum_{n=0}^\infty g(n)\frac{x^n}{n!} \label{fgr}
\end{eqnarray}
{\it Example 1:} The exponential generating function corresponding to Figure 1 is obtained by putting $L_m=0$ for
$m\neq 1$ and $V_s=1$ for all $s$. This immediately gives the  exponential generating function for the Bell
numbers $B_{1,1}(n)\equiv B(n)$ as
$\exp(\exp(x-1))$, a result we have already obtained through use of the Connected Graph Theorem above.\\
{\it Example 2:} The diagrams of Figure 1, that is corresponding to $L_m=0$ for $m\neq 1$, are in a sense generic
for the Wick contractions occurring in expressions such as $((a^{\dagger})^r
a)^n$ \cite{tobepub}. However, the
coefficients $V_s$ depend on the exponent $r$. We obtain the exponential generating function for $B_{2,1}(n)$,
corresponding to the normal ordering of $((a^{\dagger})^2 a)^n$,  by putting
$V_s=s!\ $. This immediately leads to
the formal expression
\begin{eqnarray}
G(x)&=&\left.\exp\left(x\frac{d}{dy}\right)
 \exp\left(\sum_{s=1}^{\infty} {y^s}\right)\right|_{y=0} \nonumber\\
&\equiv& \exp\left(\frac{x}{1-x}\right) \label{b21}
\end{eqnarray}
 which is  the exponential generating function for
$B_{2,1}(n)$ \cite{BPS1}, \cite{AoC}.

Many other applications and extensions of the ideas sketched in this note will be found in \cite{tobepub}.
\section*{Acknowledgements}
We thank  Carl Bender and  Itzhak Bars for interesting  discussions. PB wishes to
acknowledge support from the Polish Grant number 1 P03B 051 26.


\section*{References}
\begin{thebibliography}{99}
\bibitem{nav}Navon, A.M. 1973
Combinatorics and Fermion Algebra, {\it Nuovo Cimento} {\bf 16B}, 324
\bibitem{B} Bryant, V. 1993
{\it Aspects of combinatorics}, Cambridge Univ. Press
\bibitem{St1} Stanley, R. 1999
{\it Enumerative Combinatorics, Vol 1}, Cambridge Univ. Press
\bibitem{Leeuwen} van Leeuwen, M. 2003
personal communication.
\bibitem{BPS1} Blasiak, P., Penson, K.A. and Solomon, A.I. 2003
The general boson normal ordering problem,
{\it Phys. Lett. A} {\bf 309} 198.
\bibitem{AoC} Blasiak, P., Penson, K.A. and Solomon, A.I. 2003
The boson normal ordering problem and generalized Bell numbers,
{\it Ann. Comb.} {\bf 7} 127.
\bibitem{Katriel1} Katriel, J. 1974
Combinatorial aspects of boson algebra,
{\it Lett. Nuovo Cimento} {\bf 10} 565.
\bibitem{Katriel2} Katriel, J. 2000
Bell numbers and coherent states,
{\it Phys. Lett. A.} {\bf 273} 159.
\bibitem{Schork} Schork, M. 2003 On the combinatorics of normal
ordering of bosonic operators and deformations of it, {\it J. Phys. A:
Math. Gen.} {\bf 36} 4651
\bibitem{slo} Sloane, N.J.A. 2004 On-Line Encyclopedia of Integer Sequences,\\
available online at http://www.research.att.com/{\textasciitilde}njas/sequences/
\bibitem{Comtet} Comtet, L. 1974
\emph{Advanced Combinatorics}, Reidel,
Dordrecht.
\bibitem{tobepub} Blasiak, P., Duchamp, G., Horzela, A., Penson,
K.A. and Solomon, A.I.: Model combinatorial field theories, to be published.
\bibitem{Wilf} Wilf, H.S. 1994
\emph{Generatingfunctionology}, Academic
Press, New York.
\bibitem{Vasiliev} Vasiliev, N.A. 1998
\emph{ Functional Methods in
Quantum Field Theory and Statistical Physics}, Gordon and Breach Publishers, Amsterdam.
\bibitem{ben1} Bender, C.M, Brody, D.C. and Meister, B.K. 1999
Quantum field theory of partitions, {\it J.Math. Phys.} {\bf 40} 3239;
 Bender, C.M., Brody, D.C., and Meister B.K.  2000
 Combinatorics and field theory, {\it Twistor Newsletter} {\bf 45} 36.
\end{thebibliography}
\end{document}